\documentclass[journal]{IEEEtran}
\IEEEoverridecommandlockouts
\usepackage[dvipdf]{graphicx,color}
\usepackage{amssymb}
\usepackage{amsmath}
\usepackage{amsfonts}
\usepackage{balance}
\usepackage{multirow}
\usepackage{makecell}
\usepackage{color}
\usepackage{algorithm}
\usepackage{stfloats}
\usepackage{algpseudocode}
\usepackage{color}
\usepackage{varwidth}
\usepackage{multicol}
\usepackage{subfigure}
\usepackage{xspace}
\usepackage{enumerate}
\usepackage{xcolor,cite,etoolbox}
\usepackage{bm}
\usepackage{amsthm}
\usepackage[hyphens]{url} 

\newtheorem{observation}{Observation}
\newtheorem{proposition}{Proposition}
\newtheorem{corollary}{Corollary}
\newtheorem{remark}{Remark}
\usepackage[font=footnotesize,labelfont=scriptsize]{caption}
\usepackage{ragged2e}

\def\BibTeX{{\rm B\kern-.05em{\sc i\kern-.025em b}\kern-.08em
    T\kern-.1667em\lower.7ex\hbox{E}\kern-.125emX}}

\usepackage{titlesec}
\usepackage[margin=0.8in]{geometry}
\makeatletter 
\pretocmd\@bibitem{\color{black}\csname keycolor#1\endcsname}{}{\fail}
\newcommand\citecolor[1]{\@namedef{keycolor#1}{ \color{red}}}
\makeatother
 
\begin{document}
	
	
 {\title{\huge Finite Blocklength Performance Analysis of Uplink RSMA with Practical Imperfections}}
	\author{{Farjam Karim, \textit{Graduate Student Member}, \textit{IEEE},  Nurul Huda Mahmood, \textit{Member}, \textit{IEEE}, \\ Arthur Sousa de Sena, \textit{Member}, \textit{IEEE}, Deepak Kumar, \textit{Member}, \textit{IEEE}, 
      Matti Latva-aho, \textit{Fellow}, \textit{IEEE}}\\
\thanks{\hrulefill}
		\thanks{F. Karim, N.H. Mahmood, and M.Latva-aho are with Centre for Wireless Communications, University of Oulu, Finland. Email: \{farjam.karim, nurulhuda.mahmood,  matti.latva-aho\}@oulu.fi.\\
       A.S de Sena is with Ericsson, Sweden. (e-mail: arthurssena@ieee.org).\\
        D. Kumar is with Department of Electronics and Communication Engineering, Motilal Nehru National Institute of Technology Allahabad, Prayagraj, 211004, India (email: dkumar@mnnit.ac.in).
 } 
 \thanks{Copyright (c) 20xx IEEE. Personal use of this material is permitted. However, permission to use this material for any other purposes must be obtained from the IEEE by sending a request to pubs-permissions@ieee.org}
 \thanks{This research was supported by the Research Council of Finland (former Academy of Finland) through the 6G Flagship  (Grant Number: 369116), Business Finland’s 6GBridge- 6CORE Project (grant 8410/31/2022) and 6GBridge-Local 6G project (Grant No. 8002/31/2022), Tauno Tönningin säätiö grant and Nokia Foundation Doctoral Scholarship.}
 
 }
	%
	\maketitle
\begin{abstract}
Future generations of wireless networks will support massive communication and hyper-reliable low-latency communications, which are fundamental to large-scale Internet of Things (IoT) deployments with stringent performance requirements. To support such diverse and demanding requirements, robust and flexible multiple access schemes are essential. In this work, we develop a comprehensive analytical framework for evaluating uplink rate-splitting multiple access (RSMA) in the finite blocklength regime under realistic non-ideal conditions. Specifically, we derive closed-form expressions for the block error rate (BLER) and goodput, accounting for imperfect channel state information at the receiver (CSIR), imperfect successive interference cancellation (SIC), and hardware impairments. To reflect practical wireless environments, we model the channel using Nakagami-$m$ fading model. Our analysis shows that at lower transmit power levels, imperfect CSIR is the dominant performance-limiting factor, whereas hardware impairments and imperfect SIC become more significant in degrading system performance as the transmit power increases. To further understand the system behavior, we derive asymptotic BLER expressions. We also analyze the impact of rate allocation, demonstrating its critical role in ensuring fairness and reliability. Monte Carlo simulations validate our analytical expressions. Notably, our results show that uplink RSMA significantly outperforms NOMA even when RSMA operates under harsh imperfect CSIR. This superiority establishes RSMA as a robust and efficient candidate for future wireless networks in practical, constrained environments.
	\end{abstract}

\section{Introduction}
\IEEEPARstart{T}{he} advent of sixth-generation (6G) wireless communication is expected to revolutionize connectivity by addressing the growing demands for higher throughput, ultra-reliable communication, low latency, andmassive device connectivity driven largely by the proliferation of Internet of Things (IoT) applications~\cite{zhang_jiaqi}. While current wireless networks struggle to support the increasing density of connected devices, 6G is expected to overcome these limitations. It is anticipated to significantly enhance data rates, reliability, and quality of experience across diverse applications, including immersive communication, hyper-reliable and low-latency communication (HRLLC), and massive machine-type communication (mMTC)~\cite{Matti_2019, bruno_tut_24}.

These advancements are particularly critical for industrial IoT applications. In mMTC scenarios such as industrial automation, healthcare, and autonomous vehicles, ensuring reliable data transmission is paramount. The decoding error probability must be extremely low to guarantee accurate message delivery. Additionally, mMTC requires ultra-low latency and consistent Quality of Service (QoS) throughout the communication process, balancing throughput, reliability, and delay constraints. Traditional communication models, such as the Shannon capacity formula, assume long or infinite blocklengths (IFBL), where error correction techniques can be highly effective. However, mMTC often relies on short-packet transmissions, where conventional models fail to provide accurate reliability guarantees.

To tackle this issue, researchers have introduced finite blocklength (FBL) transmission theory~\cite{Polyanski_channelcoding_10May}. FBL offers a theoretical foundation for understanding the interplay between blocklength, error probability, and data rate. By exploring FBL principles, researchers can design coding strategies that reduce error probability in short-packet transmissions, thereby ensuring the high reliability required for mMTC applications~\cite{Onel_statisticaltools_proceedings_Nov23}.

As wireless networks evolve, optimizing limited radio resources becomes increasingly important, particularly for FBL transmissions. Traditional orthogonal multiple access (OMA) techniques, such as time division multiple access (TDMA) and orthogonal frequency-division multiple access (OFDMA), face limitations in handling massive access demands. This is because OMA schemes allocate dedicated, non-overlapping resources to each user, which leads to inefficient resource utilization and scalability issues when the number of users grows. To overcome these challenges, advanced multiple access techniques like non-orthogonal multiple access (NOMA) and rate-splitting multiple access (RSMA) have emerged~\cite{Bruno_NOMAefficent_OjCom_Jun21, Ghosh_comaprision_NOMA_OMA_Access_Jan23, Bruno_proc, yuan_lei_iot_2025, ngoc_phuc_tvt_2022, sutanu_ghosh_tccn_2025}. These techniques enhance spectral efficiency, interference management, and overall system performance, making them promising solutions for 6G wireless networks. Although the idea of rate-splitting as a multiple access strategy dates back to early work in 1996~\cite{Rimoldi_TII_96}, the topic remained relatively underexplored for many years. More recently, however, RSMA has gained significant attention, demonstrating not only improved spectral efficiency but also greater robustness to imperfect channel state information (CSI), thus, making it more attractive for interference-limited environments~\cite{yijie_comm_survey_22, Bruno_Jsac, Karim_TWC_25}.

\subsection{RSMA Working Principle}
This section briefly overviews RSMA principles to provide a clear understanding of its operation in both downlink and uplink communication scenarios. 
In the downlink, the transmission process begins with the transmitter splitting each user’s message into a shared (common) part and user-specific (private) part. The common parts of all users are aggregated into a single common stream intended for every receiver, while the private part are encoded into separate private streams. These are overlaid using superposition coding to form one superimposed signal. Upon reception, all users first attempt to decode the common stream using a universally accessible codebook. During this step, all the private streams are treated as noise. Once the common stream is extracted, individual users apply successive interference cancellation (SIC) to decode their respective private messages~\cite{Karim_TWC_25, Bruno_Jsac, Bruno_NOMAefficent_OjCom_Jun21}.

In the uplink scenario, a different strategy is employed. Except for the user with the weakest channel condition, each user splits its message and encodes into two or more parts, generating multiple streams. All streams are then transmitted simultaneously to the receiver, which employs SIC to sequentially decode and recover the original messages~\cite{yliu_twc_24_june, karim_wcl_uplink_25, Jiawei_clercx_Mar_CL23}. During system initialization, the receiver determines which users should perform message splitting and allocates power accordingly across their streams. This strategy allows RSMA to fully exploit the capacity region without relying on time-sharing. Such mechanisms make RSMA a highly flexible and robust multiple access strategy, particularly in dynamic wireless environments with fading and interference.

\subsection{Related Works} The performance of downlink RSMA has been extensively studied in the literature both in IFBL and FBL regime~\cite{Karim_TWC_25, farjam_wcnc_24, sandeep_twc_23, Allu_tccn_2025, zhu_iot_24, bruno_tut_24}. For instance, the work in~\cite{Karim_TWC_25} investigated simultaneous wireless information and power transfer (SWIPT)-enabled-RSMA in FBL regime. The authors derived key performance metrics such as block-error-rate (BLER) and goodput considering imperfect channel state information (CSI), imperfect SIC and hardware impairments at the transceiver. Similarly, in~\cite{farjam_wcnc_24}, the authors have derived expressions for outage probability (OP), throughput and total energy harvested considering linear and non-linear characteristic of the circuit for SWIPT-aided downlink RSMA network at IFBL. Reference~\cite{sandeep_twc_23}, studied RSMA for a unmanned-aerial vehicle (UAV)-assisted downlink network under both IFBL and FBL transmission. The study derived closed-form expressions for key performance metrics, and proposed a joint UAV positioning and resource allocation algorithm to maximize the ergodic sum rate under FBL constraints, considering imperfect CSI and SIC. In~\cite{zhu_iot_24}, the authors have formulated a joint optimization problem for beamforming, user scheduling and rate allocation under FBL regime. The authors in~\cite{Allu_tccn_2025} investigate an full-duplex RSMA system to improve the spectral efficiency and energy efficiency trade-off under imperfect CSI and inter-user interference. Lastly, in~\cite{bruno_tut_24}, the authors have provided a detailed review of multiple access techniques for 6G wireless networks.

While downlink RSMA has been extensively studied, investigations into uplink RSMA remain limited, with most existing works focusing on idealized conditions such as perfect CSI, perfect SIC, and no hardware impairments or analyzing performance exclusively under IFBL assumptions~\cite{yliu_twc_24_june, karim_wcl_uplink_25, xilai_twc_2026, khisa_tvt_2024}. For example, in~\cite{yliu_twc_24_june}, the authors provided an analytical study of uplink RSMA combined with hybrid automatic repeat request for IFBL, aiming to enhance transmission reliability. In~\cite{karim_wcl_uplink_25}, the authors derived OP, asymptotic OP, and throughput expressions considering imperfect CSI and imperfect SIC for IFBL transmissions. In~\cite{xilai_twc_2026}, RSMA in cell-free massive multiple-input multiple-output systems is studied under hardware impairments and imperfect CSI, where closed-form spectral efficiency expressions are derived. The authors in~\cite{khisa_tvt_2024} investigate a cooperative uplink RSMA framework, focusing on joint beamforming and power allocation to maximize the minimum signal-to-interference-plus-noise ratio (SINR).

In the context of FBL communications, the available literature on uplink RSMA remains limited and is mostly restricted to specific scenarios and simplified system assumptions~\cite{Jiawei_clercx_Mar_CL23, yixin_gc_2023, yuang_twc_2025, mehmet_tcom_jan_25}. For instance, in~\cite{Jiawei_clercx_Mar_CL23}, the authors investigated the performance of uplink RSMA in the FBL regime under ideal assumptions. In~\cite{yixin_gc_2023}, uplink RSMA for ultra-reliable low-latency communication (URLLC) is analyzed, and a blocklength minimization problem is formulated under reliability and throughput constraints. In~\cite{yuang_twc_2025}, a stochastic network calculus-based framework is developed for RSMA-assisted uplink xURLLC systems under imperfect CSI and FBL regimes, focusing on QoS provisioning through short-packet size and power optimization. Finally,~\cite{mehmet_tcom_jan_25} investigates uplink RSMA systems in the FBL regime in the context of data-oriented communication, focusing on delay-outage rate analysis under rate-adaptive and power-adaptive transmission schemes considering imperfect CSI only.

\subsection{Motivation and Contributions}
Among candidate multiple access schemes for massive IoT, OMA allocates orthogonal resources to individual devices and therefore becomes increasingly resource-inefficient as the number of devices grows, posing a challenge for massive connectivity. NOMA improves spectral efficiency by allowing users to share the same resources and relying on SIC at the receiver; however, its performance can become sensitive to the degree of channel or power disparity among users, since each user's message is represented by a single stream occupying one position in the SIC decoding order. This can limit its effectiveness when multiple devices experience similar propagation conditions, which can arise in dense IoT deployments. As discussed in Section~I-A, uplink RSMA can alleviate this limitation by allowing users to split their messages into multiple streams, providing additional flexibility in SIC decoding and rate allocation and reducing the reliance on strong channel disparities. This flexibility makes uplink RSMA attractive for massive IoT scenarios, where devices may experience heterogeneous or similar channel conditions and where short-packet transmission and low-cost hardware impose additional reliability constraints. In practical deployments, devices can be scheduled or grouped into small transmission clusters to control receiver complexity, while the message-splitting structure of uplink RSMA can be applied within each cluster. These considerations motivate our study of uplink RSMA under practical short-packet transmission conditions. 

To the best of our knowledge, an analytical framework for uplink RSMA networks operating in the FBL regime under practical non-ideal conditions remains unaddressed in
existing literature. While recent works have investigated downlink RSMA in FBL, directly extending those methods to the uplink is not straightforward because the two domains differ fundamentally. In the downlink, message splitting and superposition coding are centralized at the transmitter, whereas in the uplink, message splitting is distributed across user(s) and stream decoding is centralized at the receiver. This asymmetry introduces unique challenges in terms of allocation, interference management, and error propagation under FBL, which existing downlink analyses cannot capture. This work fills that gap by investigating the reliability of uplink RSMA in IoT-driven 6G scenarios, such as industrial automation and mMTC, where HRLLC is imperative. By incorporating imperfect CSIR, imperfect
SIC, hardware impairments, and Nakagami-$m$ fading, we model the realistic challenges these future systems are
expected to face. Importantly, this work aligns with the ongoing vision of 6G standardization.  The 3GPP 6G Workshop held in March 2025 in South Korea emphasized resiliency and efficiency as key enablers for future networks~\cite{3Gpp}, as well as the need for simple, ubiquitous, and resilient system designs. Our framework directly supports this by quantifying performance limits and trade-offs under practical constraints. Furthermore, with the European Telecommunications Standards Institute   Industry Specification group  on multiple access techniques recognizing RSMA as a strong candidate for next-generation multiple access~\cite{etsi2024}, our focus on the uplink particularly under non-ideal, real-world conditions offers a timely and foundational contribution.  This positions our work to not only inform academic discourse but also to support the recently initiated 6G standardization efforts.
The key contributions of our work are summarized as follows:

\begin{itemize}
    {\item We develop the first analytical reliability framework for uplink RSMA in the FBL regime under joint practical impairments, including imperfect CSIR, imperfect SIC, and hardware distortions. Closed-form expressions for BLER and goodput are derived, enabling tractable performance evaluation beyond idealized assumptions. The developed framework is further benchmarked against the ideal case (perfect CSIR, perfect SIC, and no hardware impairments) to quantify performance degradation.
    \item We derive asymptotic BLER expressions that reveal the emergence of a non-zero error floor at high transmit power. Our analysis explicitly characterizes how this floor is jointly governed by imperfect SIC and hardware impairments, thereby establishing fundamental reliability limits of uplink RSMA. 
    \item We validate the analytical results via Monte Carlo simulations and provide key design insights by quantifying the trade-off between blocklength and transmission rate. Specifically, we show that increasing blocklength improves reliability while reducing the inherent advantages of FBL. We further analyze the impact of rate allocation, highlighting its critical role in uplink RSMA performance.
    \item We demonstrate that uplink RSMA outperforms NOMA even under more severe CSIR imperfections, underscoring its robustness in practical scenarios. This establishes RSMA as a resilient multiple access strategy for mMTC and emerging 6G IoT applications.}
\end{itemize}
\subsection{Notations:}
Let $f_{X}(x)$ and $F_{X}(x)$ denote the probability density function (PDF) and cumulative distribution function (CDF) of a random variable $X$, respectively. The Nakagami-$m$ fading model is characterized by the parameters $m$ and $\hat{\Omega}$, which describe the shape and the average power of the fading. The notation $\mathcal{CN}(0,\sigma^2)$ represents a circularly symmetric complex Gaussian distribution with zero mean and variance $\sigma^2$ and $Q(\cdot)$ denotes the Gaussian Q function. The factorial of a non-negative integer $n$ is denoted by $n!$, and $\exp(\cdot)$ is the exponential function. Furthermore, $\binom{\cdot}{\cdot}$ denotes the binomial coefficient, while $\gamma(\cdot, \cdot)$ and $\Gamma(\cdot)$ represent the lower incomplete gamma function and the complete Gamma function, respectively.

\section{System Model}\label{sec:system_model}
  
\begin{table}				\renewcommand{\arraystretch}{1.5}
		\centering
		\caption{ Key Variables.}
		\label{t2}
		\begin{tabular}{|l|p{5.3cm}|}
			\hline
			\textbf{Symbol}         & \textbf{Definition}           \\ \hline
			
			$P_i$  &     Transmit power of $U_i\; \forall i\{1,2\}$ 	 \\ \hline

   ${h}_i$   &     Estimated channel gain    \\ \hline

${h}_{ie}$   &     Channel estimation error    \\ \hline
   
			$m_{i}$   &     Shape parameter  \\ \hline
			
			$ \hat{\Omega}_{i} \hspace{0.3em} 
 \text{and} \hspace{0.3em} \Omega_{i}    $   &     Spread parameter under  estimated CSIR and perfect case  \\ \hline

   $\Omega_{h_{ie}}$   &    Variance of channel estimation error    \\ \hline

   $\chi_{21}\hspace{0.2em}\text{and}\hspace{0.2em} \chi_{22}$ & $U_2$'s first and second message component\\ \hline
			
			$\varkappa_{21}\hspace{0.2em}\text{and}\hspace{0.2em} \varkappa_{22}$ & Power allocation for $\chi_{21}$ and $\chi_{22}$   	  \\ \hline

   ${\lambda_i}$       & Parameter indicating the quality of channel estimation 	 	 \\ \hline

    $\rho_i$       & SNR 	 	 \\ \hline

   $\eta_{ti}\hspace{0.2em} \text{and}\hspace{0.2em}  \eta_{r}$   &   Level of hardware impairment at the transmitter and receiver side  \\ \hline

			$\alpha_i$       & SIC imperfection level \\ \hline
  
   $\psi_{(\cdot)}$    & Blocklength\\ \hline
			$\vartheta_{i}$       & $U_1$ and $U_2$ BLER expression for $U_i$	 \\ \hline

  $\aleph_{i}$       &    Asymptotic BLER expression for $U_i$ 	 \\ \hline

   $\mathcal{G}_{i}$  &  Goodput expression for $U_i$\\ \hline

    $\tau$ \text{and}  $\varpi$  &  Path-loss exponent and reference path-loss\\ \hline
  $\varepsilon_{(\cdot)}$      & Channel uses for training phase \\ \hline
		$d_{i}$    & Distance between $U_i$ and the receiver\\ \hline
		\end{tabular}
	\end{table}

\begin{figure}[t]
    \centering
    \begin{minipage}[b]{0.4\textwidth}
        \centering
        \includegraphics[width=\textwidth]{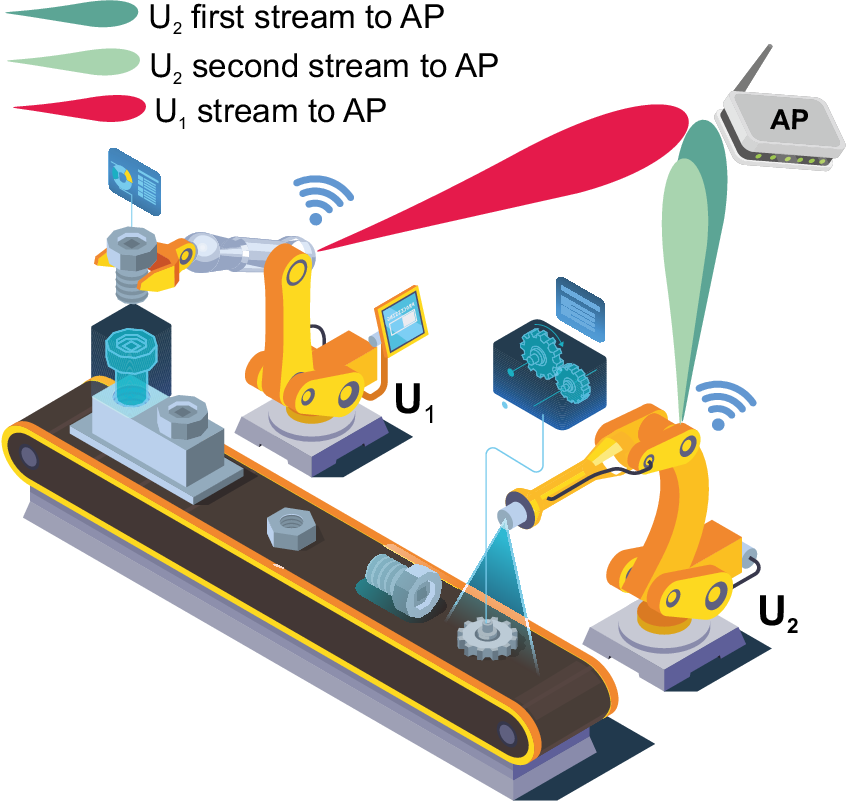}
        \caption{Uplink RSMA-aided network in an mMTC scenario.}
        \label{system_model}
    \end{minipage}
\end{figure}

We consider an RSMA-aided uplink wireless system where two users\footnote{This model can be extended to multi-user uplink scenarios by grouping users into pairs. Such pairs may be scheduled over orthogonal time or frequency resources, allowing the proposed analysis to be applied independently to each pair.}, $U_1$ and $U_2$, transmit data to an access point (AP) as illustrated in Fig.~\ref{system_model}. Each device is equipped with a single antenna\footnote{ Future work could explore extensions to multiple antennas at either end, but this is beyond the scope of the current study.}. The channel gain between user $U_i$, for $i \in \{1,2\}$, and the AP is expressed as $h_i = \hat{h}_i + h_{ie}$, where $\hat{h}_i$ denotes the estimated channel gain, while $h_{ie}$ accounts for the channel estimation error. The error term is modeled as~$h_{ie} \sim \mathcal{CN}(0, \Omega_{h_{ie}})$.  
The estimated channel gain $\hat{h}_i$ follows a Nakagami-$m$ distribution. According to~\cite{Deepak_SWIPT_SystemJ_23Mar, lei_wang_wcl_2012, narasimhan_gc_2006, singya_cl_2018}, the variance of the channel estimation error is given by  
$\Omega_{h_{ie}} = \frac{\Omega_i}{1 + \lambda_i \rho_i \Omega_i}$,  
where $\Omega_i$ represents the total variance of the channel gain, $\lambda_i > 0$ is determined by the cost of CSI acquisition, in terms of pilot power consumption, and characterizes the quality of channel estimation~\cite{lei_wang_wcl_2012}, and $\rho_i = P_i / \sigma^2$ is the signal-to-noise ratio (SNR). Consequently, the variance of the estimated channel component is calculated as $\hat{\Omega}_i = \Omega_i - \Omega_{h_{ie}}$.    
In this setup, $U_2$ is assumed to experience more favorable channel conditions, allowing it to split its message $\chi_2$ into two components, $\chi_{21}$ and $\chi_{22}$. The  SNR $\rho_2$ is allocated between these components as $\varkappa_{21}$ and $\varkappa_{22}$, ensuring that $\varkappa_{21} + \varkappa_{22} = 1$. Meanwhile, $U_1$ transmits its message $\chi_1$ without splitting, utilizing its full SNR. The messages are encoded into streams $s_{21}$, $s_{22}$, and $s_1$, which are transmitted concurrently to the AP. The received signal at the AP is thus, given by 
\begin{align}
    y&= \left(\hat{h}_1+h_{1e}\right)\left(\sqrt{P_1}s_1+\kappa_{t1}+\kappa_{r}\right)+\left(\hat{h}_2+h_{2e}\right)\nonumber\\
&\times\left(\sqrt{P_2\varkappa_{21}}s_{21}+\sqrt{P_2\varkappa_{22}}s_{22}+\kappa_{t2}+\kappa_{r}\right)+\mathfrak{n}_a,
\end{align}
where $\kappa_{ti}\sim \mathcal{CN}\left(0, P_i\eta^2_{ti}\right)$ and $\kappa_{r}\sim \mathcal{CN}\left(0, P_i\eta^2_{r}\right)$ denote the distortion noises, with $\eta_{ti}$ and $ \eta_{r}$ being the level of
hardware impairments\footnote{Wireless transceivers inevitably experience hardware limitations, including phase instabilities, mismatches between I/Q components, and nonlinear distortions introduced by power amplifiers. While mitigation techniques can reduce their impact, residual impairments always remain and continue to degrade overall system performance~\cite{tim_book_RF, nikolaos_tvt_2017, Holma_book}.} at the $U_i$ and the AP side, respectively, $\forall i\in\{1, 2\}$, and $\mathfrak{n}_{a}\sim\mathcal{CN}\left(0, \sigma^2\right)$ is the additive white Gaussian noise (AWGN) at the receiver. We assume that $s_{21}$ is the first stream to be decoded. Based on~\cite{karim_wcl_uplink_25}, the optimal decoding order to cover the entire capacity region is $s_{21} \to s_{1} \to s_{22}$. If any stream cannot be decoded, the process stops, preventing the decoding of subsequent streams. Thus, to fully recover the streams from $U_2$, the AP must decode all three streams via SIC. Consequently, the SINR for $U_2$'s $s_{21}$ message is given by
\begin{align}\label{U_22_SINR}
    \Upsilon_{21} &\!=\! \frac{
    \left|\hat{h}_2\right|^2\rho_2\varkappa_{21}
    }{
    \begin{aligned}
    &\left|\hat{h}_2\right|^2\!\!\rho_2\left(\varkappa_{22}\!+\! \eta^2_{t2} \!+\! \eta^2_{r}\right)
   \! \!+\! \left|\hat{h}_1\right|^2\!\rho_1\left(1 \!+\! \eta^2_{t1} \!+\! \eta^2_{r}\right) \\
    &\!+\! \rho_1\Omega_{h1e}\!\left(1 \!+\! \eta^2_{t1} \!+\! \eta^2_{r}\right) \!\!+\! \rho_2\Omega_{h2e}\!\left(1 \!\!+\! \eta^2_{t2} \!+\! \eta^2_{r}\right)\! \!+\! 1
    \end{aligned}
    }.
\end{align}
For $U_1$, the AP is required to decode only the streams $s_{21}$ and $s_1$. Once $s_{21}$ is successfully decoded, the AP will then decode $s_1$ with the SINR of
\begin{align}\label{U_1_SINR}
    &\Upsilon_{1} \!=\! \frac{
    \left|\hat{h}_1\right|^2\rho_1
    }{
    \begin{aligned}&
    \left|\hat{h}_2\right|^2\!\!\rho_2\!\left(\Xi_2 \!+\! \eta^2_{t2} + \eta^2_{r}\!\right) \!+\! \left|\hat{h}_1\right|^2\!\!\rho_1\!\left(\eta^2_{t1} \!+\! \eta^2_{r}\right)\!+\!1 \\
    &+ \!\rho_1\Omega_{h1e}\!\left(\!1 \!+\! \eta^2_{t1} \!+\! \eta^2_{r}\!\right)\!+\! \rho_2\Omega_{h2e}\!\left(\Xi_2 \!+\! \eta^2_{t2} \!+\! \eta^2_{r}\right) 
    \end{aligned}
    },
\end{align}
where $\Xi_2 = \left(\varkappa_{22}+\alpha_2\varkappa_{21}\right)$ and $\alpha_2$ denotes the imperfect SIC factor after decoding \eqref{U_22_SINR}, with $\alpha_2 \in \left[0, 1\right]$~\cite{farjam_wcnc_24, vu_hoc_iot_22}.  Once both streams are decoded successfully, the AP attempts to decode the $s_{22}$ stream with a SINR of
\begin{align}\label{U_22_final_SINR}
     &\Upsilon_{22} \!=\! \frac{
    \left|\hat{h}_2\right|^2\rho_2\varkappa_{22}
    }{
    \begin{aligned}&
    \left|\hat{h}_2\right|^2\!\!\rho_2\!\left(\alpha_2\varkappa_{21}\! \!+\! \eta^2_{t2} \!+\! \eta^2_{r}\right) \!\!+\! \!\left|\hat{h}_1\right|^2\!\!\rho_1\!\left(\!\alpha_1\! \!+\! \eta^2_{t1} \!+\! \eta^2_{r}\right)\!\!+\!\!1 \\
    &+ \rho_1\Omega_{h1e}\!\left(\!\alpha_1 \!+\! \eta^2_{t1} \!+\! \eta^2_{r}\!\right)\!+\! \rho_2\Omega_{h2e}\!\left(\Xi_2 \!+\! \eta^2_{t2} \!+\! \eta^2_{r}\right) 
    \end{aligned}
    }.
\end{align}
By setting $\alpha_i =0$ , the SINR for perfect SIC can be obtained.
\section{Performance Evaluation}\label{performance}
This section investigates the performance of the uplink RSMA-aided system in the FBL regime. Note that extending the framework to more than two users is non-trivial, since each additional user introduces extra random variables and unknown decoding dependencies, which significantly increase the mathematical complexity. However, the current analysis remains applicable to larger systems when users are paired and scheduled across different time or frequency resources. In such cases, each pair can be analyzed using the proposed two-user model, while scalability to multiple pairs can be addressed at the resource scheduling level. A full extension to multi-user simultaneous transmission is left for future work. We derive closed-form expressions for the BLER, goodput, and asymptotic BLER. Given that $\hat{h}_{i}$ follows a Nakagami-$m$ distribution, the squared magnitude, $|\hat{h}_{i}|^2$, results in Gamma distribution. The PDF and CDF of $|\hat{h}_{i}|^2$ are given as~\cite{Karim_TWC_25}:
\vspace{1em}
 \begin{align}\label{pdf}
     f_{\left|\hat{h}_i\right|^2}(x)=\frac{m_i^{m_i} x^{m_i-1}}{\hat{\Omega}_i^{m_i} \Gamma\left(m_i\right)} \exp \left(-\frac{m_i}{\hat{\Omega}_i} x\right), \quad x\geq 0.\end{align}

    \begin{align}  \label{CDF} F_{\left|\hat{h}_i\right|^2}(x)= \frac{1}{\Gamma\left(m_i\right)} \gamma\left(m_i, \frac{ m_i }{\hat{\Omega}_i}x\right), \quad x\geq 0.
 \end{align}
\subsection{BLER Analysis:} The BLER of $U_{1}$ at the AP for decoding the $x_j$ streams for $\psi_j$ channel uses and $\beta_j$  bits/slot when $\psi_j > 100 $ can be approximated as~\cite{chung_shortpack_Tvt_21mar},
 \begin{align}\label{Qfunc}
     \vartheta_{1} \approx \mathbb{E}\left\{Q\left(\frac{C\left( \Upsilon_{j}\right)-r_j}{\sqrt{V\left( \Upsilon_j\right) / \psi_j}}\right)\right\},
 \end{align} where $C\left( \Upsilon_{j}\right) = \log_2 (1 + \Upsilon_j)$ represents the instantaneous channel capacity, $r_1 = \frac{\beta_1}{\psi_1}$ is the rate, and $V\left( \Upsilon_j\right)$ denotes the channel dispersion. The channel dispersion quantifies the variability of the channel compared to a deterministic channel with the same capacity and is given by $\left(1 - (1 + \Upsilon_j)^{-2}\right)(\log_2 e)^2, \; \forall j \in \{21, 1\}$. As $U_2$'s message is split into two parts, the rates for the resulting streams $ s_{21} $ and $ s_{22} $ are defined as  $r_{21} = \frac{\delta\beta_2}{\psi_2}$   and $ r_{22} = \frac{(1 - \delta) \beta_2}{\psi_2}$, respectively.
Here, $\delta$ is the rate allocation factor, with $0 \leq \delta \leq 1$,~ \cite{Jiawei_clercx_Mar_CL23, karim_wcl_uplink_25}. 

To obtain the overall BLER $U_1$, we account for the sequential decoding process at the AP. Since, $s_{21}$ is decoded before $s_{1}$, an error in decoding $s_{21}$ results in a decoding failure for $U_1$'s stream. If $s_{21}$ is successfully decoded, the residual interference after SIC, together with the CEE and hardware impairments, determines the decoding reliability of $s_{1}$. Accordingly, the overall BLER is obtained by accounting for the failure probability of the first decoding stage and the conditional failure probability of the second decoding stage, both under the FBL regime, as given in the following proposition.

\begin{proposition}\label{Bler_A}
The BLER of $U_1$'s message at the AP considering imperfect CSIR, hardware impairments and imperfect SIC can be evaluated as   
  \begin{align}\label{BLER_U_1}
      \vartheta_1 \approx \vartheta_{21}+\left(1-\vartheta_{21}\right)\vartheta_{11},
  \end{align}
  where $\vartheta_{21}$ given in \eqref{BLER_21} $ \text{and} \;\vartheta_{11}$ is given in \eqref{BLER_11} with parameters defined as follows:     
\begin{align}
    &\mu_j = 2^{r_j}-1, \quad\Lambda_j = \left({2\pi}{\left(2^{2r_j}-1\right)}/\psi_j\right)^{{-1}/{2}}, \nonumber\\ 
    &\xi_j = \mu_j - \frac{1}{2\Lambda_j}, \quad \zeta_j = \mu_j + \frac{1}{2\Lambda_j}, E_{1} =\left(\eta^2_{t1} +\eta^2_{r}\right) , \nonumber\\ 
    &D_{21}\!=\!\frac{m_2 \left(\left(\zeta_{21}+\xi_{21}\right)/2\right) }{{\hat{\Omega}_2}\rho_2\left(\varkappa_{21}\!-\!\left(\left(\zeta_{21}+\xi_{21}\right)/2\right)A_{21}\right)}, N_1=\frac{m_2}{\hat{\Omega}_2}, \nonumber\\
    & Z_{2}=\rho_1\Omega_{g1e}\left(1+\eta^2_{r}+\eta^2_{t1}\right)+\Omega_{g2e}Z_1 +1, \nonumber\\
    & A_{21}=\left(\varkappa_{22}+\eta^2_{r}+\eta^2_{t2}\right), B_{11}=\rho_1\left(1 + \eta^2_{t1} +\eta^2_{r}\right), \nonumber\\
    & C_{21}=\rho_2\Omega_{g2e}\left(1 + \eta^2_{t2} +\eta^2_{r}\right)+B_{11}\Omega_{g1e}+1,  L_{21}=\frac{m_1}{\hat{\Omega}_1},\nonumber\\
    & A_{12}= \rho_1\Omega_{g1e}\left(1 + \eta^2_{t1} + \eta^2_{r}\right), Z_1=\left(\Xi_2+\eta^2_{t2} +\eta^2_{r}\right)\rho_2,  \nonumber\\
    &\text{and}\; W_{1}=\frac{m_1 \left(\left(\zeta_{11}+\xi_{11}\right)/2\right) }{{\hat{\Omega}_1}\rho_1\left(1-\!\left(\left(\zeta_{11}+\xi_{11}\right)/2\right)E_1\right)}. \nonumber  
\end{align}
\end{proposition}
\begin{proof}
    Please refer to Appendix A.
\end{proof}

\begin{observation}\label{Rem1}
 The expression in Proposition~\ref{Bler_A} provides a compact characterization of the joint impact of FBL and system imperfections on the BLER. Specifically, $\mu_j$, $\Lambda_j$, $\xi_j$, and $\zeta_j$ capture the FBL dependence on the transmission rate and blocklength, while $A_{21}$, $B_{11}$, $C_{21}$, $Z_1$, and $Z_2$ incorporate the effects of interference, imperfect CSIR, and transmitter and receiver hardware impairments. The parameters $D_{21}$, $N_1$, $L_{21}$, and $W_1$ arise from averaging the conditional BLER over the fading distribution. This structure also provides an intuitive explanation of the high-SNR behavior. Although increasing SNR generally improves decoding reliability, residual hardware distortion and imperfect SIC, represented by $\eta_{t1}$, $\eta_{t2}$, $\eta_{r}$, and $\alpha_2$, can prevent the BLER from vanishing and lead to a non-zero error floor. Thus, the resulting BLER is governed by a trade-off between the FBL penalty and the  effects of CEE, hardware impairments, and SIC imperfections. 
\end{observation}

\begin{remark}
    The BLER of $U_1$'s message at the AP considering perfect CSIR, no hardware impairments and perfect SIC can be obtained by letting channel estimation quality parameter, $\lambda_i\to\infty$ and substituting hardware impairment terms, $\eta^2_{ti}=\eta^2_{r}=0$, and imperfect SIC factor, $\alpha_2=0,\; \forall i\in \{1,2\}$, respectively, for calculating $\vartheta_1 $ in \eqref{BLER_U_1}.
\end{remark}
Following the same FBL-based approach, the BLER of $U_2$ is obtained by accounting for the sequential decoding process described earlier. Since the decoding of $s_{22}$ follows the preceding decoding stages, the reliability of $U_2$'s message depends on the successful completion of the SIC chain. The corresponding BLER expression is given in the following proposition.
\begin{proposition}\label{BLER_b}
The BLER of $U_2$'s message at the AP considering imperfect CSIR, hardware impairments and imperfect SIC  can be evaluated as   
  \begin{align}\label{BLER_U_2}
      \vartheta_2&  \approx \vartheta_{21}+\left(1-\vartheta_{21}\right)\vartheta_{11}+\left(1-\vartheta_{21}\right)\left(1-\vartheta_{11}\right)\vartheta_{22},
  \end{align}
  where $\vartheta_{22}$ given in \eqref{BLER_22} with parameters given by:
\begin{align}
    &E_{22} = \left(\alpha_2\varkappa_{21}+\eta^2_{r}+\eta^2_{t2}\right), \Lambda_k =\left({2\pi}{\left(2^{2r_k}-1\right)}/\psi_k\right)^{-\frac{1}{2}},\nonumber\\
    &  \mu_k = 2^{r_k}-1, \zeta_k \!=\! \mu_k + \frac{1}{2\Lambda_k} \forall k\in \{21, 1, 22\}, \nonumber\\
    & C_5 =\frac{m_2 \left(\left(\zeta_{22}+\xi_{22}\right)/2\right) }{{\hat{\Omega}_2}\rho_2\left(\varkappa_{22}\!-\!\left(\left(\zeta_{22}+\xi_{22}\right)/2\right)E_{22}\right)},  \xi_k= \mu_k - \frac{1}{2\Lambda_k}, \nonumber\\
    &  E_{25} =  \left(C_{22}\Omega_{g1e} +  \Omega_{g2e}Z_{1} +1\right), \text{and}\nonumber\\ 
    &C_{22}=\rho_1\left(\alpha_1+\eta^2_{r}+\eta^2_{t1}\right).\nonumber
\end{align}

\begin{observation}
Like Observation~\ref{Rem1}, the expression in Proposition~\ref{BLER_b} incorporates the effects of the considered system imperfections through the auxiliary parameters. In particular, the terms $C_{22}$, $E_{22}$, and $Z_1$ capture the effects of hardware impairments and residual SIC interference, while $E_{25}$ accounts for imperfect CSIR. The FBL dependence is reflected through $\mu_k$, $\Lambda_k$, $\xi_k$, and $\zeta_k$. Together, these terms characterize the reliability of $U_2$ under finite blocklength and non-ideal decoding conditions.
\end{observation}

\end{proposition}
\begin{proof}
  By following a similar approach as outlined in Appendix A, the proof for $\vartheta_2$ can also be obtained. 
\end{proof}
\begin{proposition}
    The BLER of $U_2$'s message at the AP considering perfect CSIR, no hardware impairments and perfect SIC can be obtained by substituting $\vartheta_{22}=\Lambda_{22}\left(\zeta_{22}-\xi_{22}\right)\left(\frac{1}{\Gamma({m_2})}\gamma\left({m_2, \frac{m_2\left(\left(\zeta_{22}+\xi_{22}\right)/2\right)}{\hat{\Omega}_2\rho_2\varkappa_{22}}}\right)\right)$ in \eqref{BLER_U_2}.
\end{proposition}
\begin{proof}
 Please refer to Appendix B.
\end{proof}
\begin{figure*}
\begin{align}\label{BLER_21}
    \vartheta_{21}\approx \Lambda_{21}\left(\zeta_{21}-\xi_{21}\right)\Bigg[ 1-\sum\limits_{{\mathcal{I}}=0}^{{m_2}-1}\sum\limits_{{\mathcal{J}}=0}^{{\mathcal{I}}}\binom{{\mathcal{I}}}{{\mathcal{J}}}\frac{\exp\left(-D_{21}C_{21}\right)\left(D_{21}B_{11}\right)^{\mathcal{J}}  \left(D_{21}C_{21}\right)^{({\mathcal{I}}-{\mathcal{J}})}L_{21}^{m_1}
      \Gamma\left({m_1}+{\mathcal{J}}\right) }{({\mathcal{I}}!)\Gamma(m_1)\left(D_{21}B_{11}+L_{21}\right)^{\left({m_1}+{\mathcal{J}}\right)} }\Bigg].
    \end{align}\vspace{2em}\hrulefill
    \begin{align}\label{BLER_11}
    \vartheta_{11}\approx \Lambda_{11}\left(\zeta_{11}-\xi_{11}\right)\Bigg[ 1-\sum\limits_{{\mathcal{K}}=0}^{{m_1}-1}\sum\limits_{l=0}^{{\mathcal{K}}}\binom{{\mathcal{K}}}{l}\frac{ \exp\left(-W_{1}Z_{2}\right)\left(W_{1}Z_{1}\right)^l  \left(W_{1}Z_{2}\right)^{({\mathcal{K}}-l)}N_1^{m_2}
     \Gamma\left({m_2}+l\right) }{({\mathcal{K}}!)\Gamma(m_2)\left(W_{1}Z_{1}+N_1\right)^{\left({m_2}+l\right)} }\Bigg].
    \end{align}\vspace{2em}\hrulefill
     \begin{align}\label{BLER_22}
    \vartheta_{22}\approx \Lambda_{22}\left(\zeta_{22}-\xi_{22}\right)\Bigg[ 1-\sum\limits_{{\mathcal{I}}=0}^{{m_2}-1}\sum\limits_{{\mathcal{J}}=0}^{{\mathcal{I}}}\binom{{\mathcal{I}}}{{\mathcal{J}}}\frac{ \exp\left(-C_{5}E_{25}\right)\left(C_{5}E_{22}\right)^{{\mathcal{J}}}\left(C_{5}E_{25}\right)^{({\mathcal{I}}-{\mathcal{J}})}  L_{21}^{m_1}
     \Gamma\left({m_1}+{\mathcal{J}}\right) }{({\mathcal{I}}!)\Gamma(m_1)\left(C_{5}C_{22}+L_{21}\right)^{\left({m_1}+{\mathcal{J}}\right)} }\Bigg].
    \end{align}\vspace{2em}\hrulefill
\end{figure*}
\vspace{-1em}
\subsection{Goodput Analysis:} Having derived the BLER expressions for $U_1$ and $U_2$, we now turn our attention to the user goodput. Expressed in nats per channel use (npcu), the goodput represents the effective communication rate when data or frames, including both the data transmission and training phases, are successfully delivered over the network using the total channel uses~\cite{chung_shortpack_Tvt_21mar}. Therefore, the corresponding goodput expressions for $U_1$ and $U_2$ are presented in the following corollaries, respectively.
\begin{corollary}
     The goodput of $U_1$'s message at the AP under imperfect CSIR, hardware impairments and imperfect SIC  can be evaluated as
     \begin{align}\label{goodput_U1}
         \mathcal{G}_1 \approx \!\Biggl[\left(1-{\frac{1}{\Psi_{11}}}\right) r_{1}\Biggl]\left(1-{\vartheta}_{1}\right),
     \end{align}
     where $\Psi_{11} = \psi_{1}+\varepsilon_{1}$ with $\varepsilon_{1}$ denoting the channel uses for training phase. 
\end{corollary}
\begin{proof}
    Proof can be obtained by following Appendix A.
\end{proof}
\begin{remark}
    By letting  the channel estimation quality parameter $\lambda_i\to \infty$, and setting the imperfect SIC factor $\alpha_i = 0$, and the hardware impairment factors $\eta^2_{r} = \eta^2_{ti} = 0\; \forall\; i\in\{1, 2\}$,  the expression for $\mathcal{G}_{1}$ for the perfect case is obtained.
\end{remark}
\begin{corollary}
     The goodput of $U_2$'s message at the AP under imperfect CSIR, hardware impairments and imperfect SIC  can be evaluated as
     \begin{align}\label{goodput_U2}
         \mathcal{G}_2 \approx \!\Biggl[\left(1-{\frac{1}{\Psi_{2}}}\right) r_{21} + \left(1-{\frac{1}{\Psi_{2}}}\right) r_{22}\Biggl]\left(1-{\vartheta}_{2}\right),
     \end{align}
     where $\Psi_{2} = \psi_{2}+\varepsilon_{2}$  with $\varepsilon_{2}$ denoting the channel uses for training phase. 
\end{corollary}
\begin{remark}
    By applying the methods outlined in Remark 2 for calculating $\vartheta_{22}$,  the expression for $\mathcal{G}_{2}$ for the perfect CSIR and perfect SIC case with ideal hardware impairments can be obtained. Note that the term ideal hardware impairments refers to  $\eta^2_{ti}=\eta^2_{r}=0\; \forall i\in\{1,2\}$.
\end{remark}
The expressions in \eqref{goodput_U1} and \eqref{goodput_U2} show that the goodput is jointly determined by the coding rate, training overhead, and BLER. Thus, increasing the transmission rate does not necessarily improve the goodput if the resulting increase in BLER outweighs the gain in transmission rate.
\subsection{Asymptotic BLER Analysis:}
The previous discussion provides analytical insights into how the BLER and goodput depends on various system imperfections and finite SNR.  Building on this, we can further investigate the system behavior by letting $\rho_i \to \infty$ to obtain the asymptotic BLER expressions for $U_1$ and $U_2$. 
This asymptotic analysis allows us to quantify the limiting performance and to corroborate the trends observed in the finite-SNR regime, thereby providing additional insights for system design. The respective asymptotic BLER for $U_1$ is evaluated in the following proposition.
\begin{proposition} The BLER of $U_1$'s message at the AP considering hardware impairments and imperfect SIC at high SNR can be evaluated as   
  \begin{align}\label{Asymp_BLER_U_1}
      \aleph_1 \approx \mathfrak{T}_{21}+\left(1-\mathfrak{T}_{21}\right)\mathfrak{T}_{11},
  \end{align}
  with the parameters given by:
  \begin{align}
      &\mathfrak{D}_{21}\!=\!\frac{m_2\hat{B}_{11}\left(\left(\zeta_{21}\!+\!\xi_{21}\right)/2\right)}{\hat{\Omega}_2\left(\varkappa_{21}\!-\!\left(\left(\zeta_{21}\!+\!\xi_{21}\right)/2\right)A_{21}\right)}, \mathfrak{T}_{21}\!=\!\Lambda_{21}\left(\zeta_{21}-\xi_{21}\right)\nonumber\\
      &\times\Bigg[1-\sum\limits^{{m_2}-1}_{{\mathcal{I}}=0}\frac{L_{21}^{m_1}{\mathfrak{D}_{21}}^{\mathcal{I}} \Gamma({m_1+{\mathcal{I}}})}{({\mathcal{I}}!)\Gamma(m_1)\left(L_{21}+\mathfrak{D}_{21}\right)^{m_1+{\mathcal{I}}}}\Bigg], \nonumber\\
& \mathfrak{Z}_1=\left(\Xi_2+\eta^2_{t2} +\eta^2_{r}\right), \hat{B}_{11}=\left(1+\eta^2_{t1}+\eta^2_{r}\right), \mathfrak{T}_{11}\!=\Lambda_{11}\nonumber\\
& \times\left(\zeta_{11}-\xi_{11}\right)\Bigg[1-\sum\limits^{{m_1}-1}_{{\mathcal{K}}=0}\frac{N_{1}^{m_2}{\mathfrak{W}_{1}}^{\mathcal{K}} \Gamma({m_2+{\mathcal{K}}})}{({\mathcal{K}}!)\Gamma(m_2)\left(N_{1}+\mathfrak{W}_{1}\right)^{m_2+{\mathcal{K}}}}\Bigg], \nonumber\\
&\text{and}\;  \mathfrak{W}_{1}=\frac{m_1\mathfrak{Z}_1\left(\left(\zeta_{11}+\xi_{11}\right)/2\right)}{\hat{\Omega}_1\left(1-\left(\left(\zeta_{11}+\xi_{11}\right)/2\right)E_{1}\right)}. \nonumber
  \end{align}
   \end{proposition}
  \begin{proof}
      Please refer to Appendix C.
  \end{proof}
  \begin{observation}
     From the asymptotic expression in~\eqref{Asymp_BLER_U_1}, it is evident that the SNR term $\rho_i$ does not explicitly appear. This implies that hardware impairments and imperfect SIC terms do not scale with the transmit power; instead, they impose a constant effect on $U_1$'s performance. Furthermore, as $\rho_i \to \infty$, the CEE terms vanish due to the formulation given in Section~\ref{sec:system_model}. Collectively, these factors influence all the parameters defined in Proposition~4. This characteristic highlights the fundamental difference between the asymptotic BLER expression~\eqref{Asymp_BLER_U_1} and the standard BLER expression~\eqref{BLER_U_1}.
  \end{observation}
  \begin{remark}
    By setting imperfect SIC term, $\alpha_i = 0$, and hardware impairments terms, $\eta^2_{r} = \eta^2_{ti} = 0\; \forall\; i\in\{1, 2\}$,  the expression for $U_1$'s asymptotic BLER ($\aleph_{1}$) for the perfect SIC case and ideal hardware impairment can be obtained. 
\end{remark}
Having evaluated the numerical expression for the asymptotic BLER of
$U_1$, we now derive the corresponding asymptotic BLER expression for $U_2$ as stated in the following proposition.
\begin{proposition} The BLER of $U_2$'s message at the AP considering hardware impairments and imperfect SIC at high SNR can be evaluated as   
  \begin{align}\label{Asymp_BLER_U2}
      \aleph_2 \approx \mathfrak{T}_{21}\!+\!\left(1-\mathfrak{T}_{21}\right)\mathfrak{T}_{11}\!+\!\left(1-\mathfrak{T}_{21}\right)\left(1-\mathfrak{T}_{11}\right)\mathfrak{T}_{22},
  \end{align}
 with the parameters given by:
  \begin{align}
      &\mathfrak{C}_{5}=\frac{m_2\mathfrak{C}_{22}\left(\left(\zeta_{22}+\xi_{22}\right)/2\right)}{\hat{\Omega}_2\left(\varkappa_{22}-\left(\left(\zeta_{22}+\xi_{22}\right)/2\right)E_{22}\right)}, \mathfrak{T}_{22}=\Lambda_{22}\nonumber\\
      & \times\left(\zeta_{22}-\xi_{22}\right)\Bigg[1-\sum\limits^{{m_2}-1}_{{\mathcal{I}}=0}\frac{L_{21}^{m_1}{\mathfrak{C}_{5}}^{\mathcal{I}} \Gamma({m_1+{\mathcal{I}}})}{({\mathcal{I}}!)\Gamma(m_1)\left(L_{21}+\mathfrak{C}_{5}\right)^{m_1+{\mathcal{I}}}}\Bigg],\nonumber\\
      & \; \text{and}\;  \mathfrak{C}_{22}=\left(\alpha_1+\eta^2_{r}+\eta^2_{t1}\right).\nonumber
  \end{align}
   \end{proposition}
  \begin{proof}
 The proof can be obtained by following a similar approach outlined in Appendix C. 
\end{proof}
  
\begin{observation}
    Similar to Observation~3, it can be observed here that the key distinction lies in the asymptotic BLER of $U_2$ compared to its standard BLER. 
\end{observation}
 \begin{proposition}
    The asymptotic BLER of $U_2$'s message at the AP considering ideal hardware impairments and perfect SIC can be obtained by substituting $\mathfrak{T}_{22}=\Lambda_{22}\left(\zeta_{22}-\xi_{22}\right)\Bigg[\frac{1}{\Gamma({m_2})\; m_2}\left({ \frac{m_2\left(\left(\zeta_{22}+\xi_{22}\right)/2\right)}{\hat{\Omega}_2\rho_2\varkappa_{22}}}\right)^{m_2}\Bigg]$ in \eqref{Asymp_BLER_U2}.
\end{proposition}
\begin{proof}
 Please refer to Appendix D.
\end{proof}
The above asymptotic expressions characterize the limiting BLER of both users when the SNR grows without bound. Together with the \eqref{BLER_U_1} and  \eqref{BLER_U_2}, they provide a direct characterization of the reliability behavior across the SNR range and identify the performance limit imposed by the considered system imperfections.
    
\section{Results and Discussions}
This section presents Monte Carlo simulation results to validate the derived closed-form expressions. The wireless link between each user $U_i$ and the AP is characterized using a distance-based path-loss model, given by $\varpi/d_i^{\tau}$, where $i \in \{1,2\}$, $\varpi = 1$~m is the reference path-loss parameter, $d_i$ is the distance from $U_i$ to the AP, and $\tau = 3.8$ is the path-loss exponent.  
    The channel estimation error variance is simulated using the formula outlined in Section~\ref{sec:system_model}, and the BLER performance is simulated using the Q-function-based expression introduced in Section~\ref{performance}. Unless otherwise stated, the simulation parameters are listed in Table~\ref{t3}. These simulation results offer key insights into the practical factors influencing system performance in the FBL regime and also serve to benchmark the accuracy of our analytical framework.
    \begin{table}[t]	\renewcommand{\arraystretch}{1.0}
		\centering
		\caption{ Simulation Parameters.}
		\label{t3}
			\resizebox{\columnwidth}{!}{\begin{tabular}{|l|l|l|l|l|l|}
			\hline
			Parameter         & Value         & Parameter & Value  & Parameter & Value  \\ \hline
			$m_1  $   &     $3 $   	&  $m_2$  &     $4$ & $\sigma^2$   &     $-100 $ dB \\ \hline
			
			 $\beta_1$   &     $50$ &	   $ \beta_{2}$    &    $ 50 $ &  $\delta $  &     $0.15$  	 \\ \hline
			
		 $\varkappa_{21}$       & $0.65 $ 	&  $\varkappa_{22} $   &     $0.35$    &      $U_1$-AP  &     $75$m  	 \\ \hline

    $U_2$-AP      & $73 $m 	&  $\psi_{1} $   &     $600 $    &      $\psi_2 $  &     $ 600$  	 \\ \hline

         $\varepsilon_{1}$      & $100 $	&   $\varepsilon_{2}$      & $100 $   &      $ $  &     $ $  	 \\ \hline 
		\end{tabular}}
	\end{table}

\begin{figure*}[t]
    \centering
    \begin{minipage}[b]{0.49\textwidth}
        \centering
        \includegraphics[width=\textwidth]{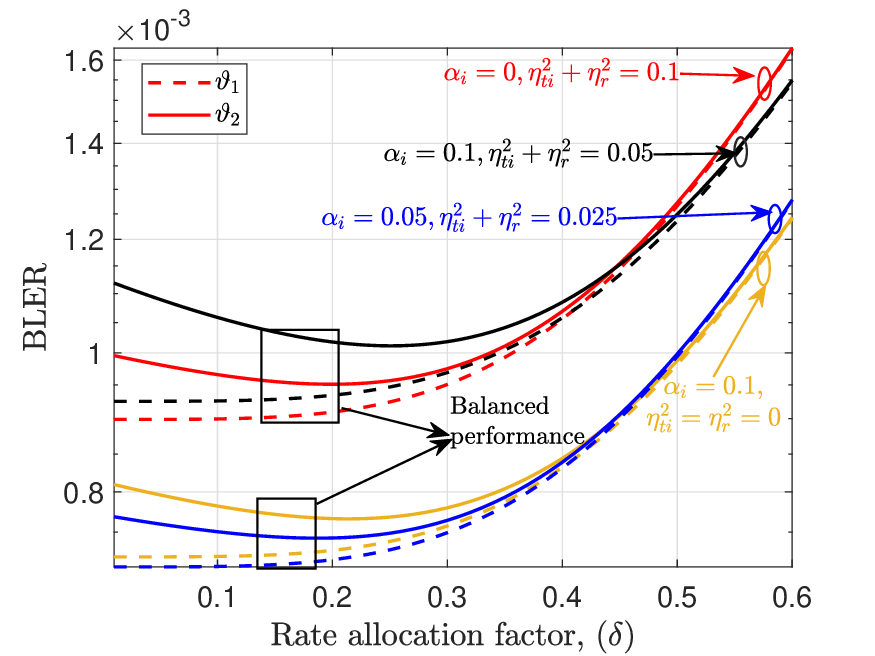}
        \caption{Effect of $\delta$ on BLER.}
        \label{rate_allocation}
    \end{minipage}
     \begin{minipage}[b]{0.49\textwidth}
        \centering
        \includegraphics[width=\textwidth]{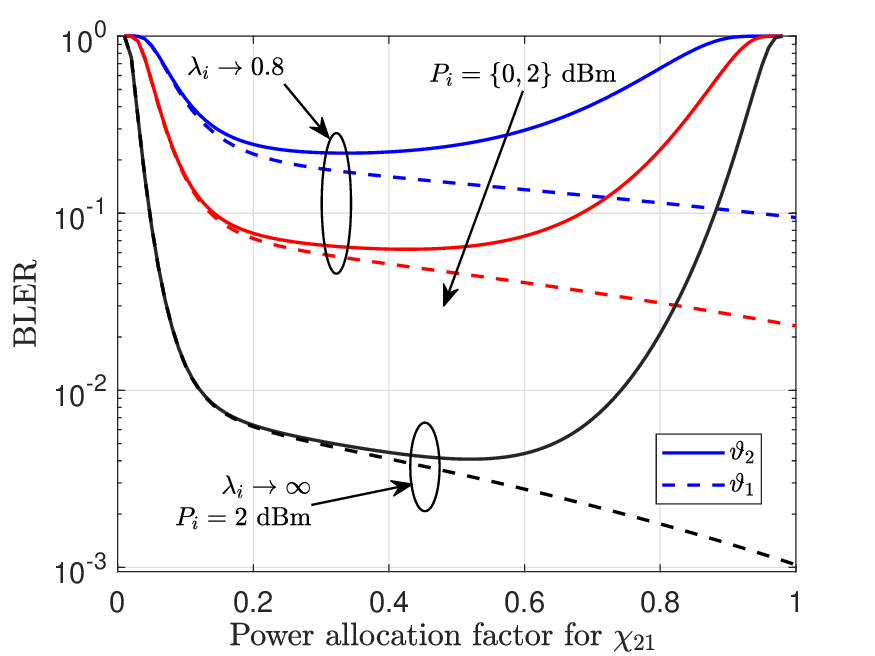}
        \caption{Effect of $\varkappa_{21}$ on BLER.}
        \label{vcv}
    \end{minipage}
\end{figure*}

Fig.~\ref{rate_allocation} illustrates the BLER performance of $U_1$ and $U_2$ as a function of the rate allocation factor $\delta$, considering channel estimation quality parameter $(\lambda_i) \to 0.7$ with various levels of imperfect SIC ($\alpha_i$) and hardware impairments ($\eta^2_{r}, \eta^2_{ti}$). The results show that a balanced BLER where both users achieve comparable reliability when $0.1 \leq \delta \leq 0.2$. Outside this range, particularly for $\delta > 0.2$, the BLER of both users deteriorates rapidly. In this regime, $U_1$ becomes increasingly dependent on $U_2$’s decoding performance, which itself degrades due to the compounded difficulty in decoding the second portion of its message.
This behavior stems from the decoding structure detailed in \eqref{U_22_SINR}, \eqref{U_1_SINR}, and \eqref{U_22_final_SINR}. As $\delta$ increases, the rate assigned to $U_2$’s second decoding message becomes severely limited, undermining the effectiveness of both message splitting and SIC. In the extreme case where $\delta = 1$, the power allocated to the second part of $U_2$’s message becomes meaningless, causing the numerator of the $r_{22}$ term to approach zero. This renders the decoding process ineffective, resulting in invalid or undefined BLER values.

It is important to clarify that \textit{balanced BLER} refers to maintaining fairness in reliability across users while meeting their respective QoS targets. A key takeaway from this figure is the distinct role of $\delta$ in uplink RSMA.  Unlike downlink RSMA, where rate allocation pertains to sharing the common rate among users $\delta$ in uplink RSMA governs how a single user's own message is split into two parts for sequential SIC decoding, thereby directly shaping the decoding dependency between users. This highlights the necessity of judiciously selecting $\delta$ to ensure system resiliency and maintain robust performance in the presence of practical impairments such as imperfect CSIR, hardware non-idealities, and SIC imperfections.  In particular, these results suggest restricting $\delta$ to the range $0.1-0.2$ as a practical guideline for maintaining balanced and reliable performance between the two users under the considered impairment levels.

Fig.~\ref{vcv} presents the BLER performance as a function of the power allocation factor $\varkappa_{21}$ associated with the first message component of $U_2$, under imperfect CSIR conditions, with $\alpha_i = 0.1$ and $\eta^2_{r} = \eta^2_{ti} = 0.05$.
The values of $\alpha_i$ and $\eta^2_{r} = \eta^2_{ti}$ are set consistent with~\cite{vu_hoc_iot_22} and \cite{nikolaos_tvt_2017}, respectively.
Several interconnected observations emerge from the figure.
First, higher channel estimation quality ($\lambda_i$) improves the BLER performance for both users. However, this improvement comes with increased interdependence between $U_1$ and $U_2$. For example, at $P_i = 2$~dBm and $\lambda_i \to \infty$, $U_1$ remains dependent on $U_2$'s decoding performance until approximately $\varkappa_{21} = 0.3$. In contrast, with a lower estimation quality of $\lambda_i \to 0.8$, this dependency threshold shifts earlier to around $\varkappa_{21} = 0.18$. While reduced dependency can enhance robustness against interference propagation, it also leads to noticeably degraded BLER for both users, highlighting a critical trade-off between performance and decoding independence.
Second, identifying the optimal value of $\varkappa_{21}$ through simulation alone proves to be non-trivial, especially given the sensitivity of BLER to variations in both $\lambda_i$ and $\varkappa_{21}$. This complexity is further amplified in dynamic scenarios. Specifically, in practical systems where users are mobile rather than stationary, continuously adapting $\varkappa_{21}$ to maintain target QoS becomes computationally intensive and energy inefficient. 
These insights underline the practical limitations of relying solely on simulation-based parameter tuning and suggest that real-time validation or adaptive learning-based mechanisms may be required for robust and resilient RSMA deployments in real-world mobile environments.

\begin{figure*}[t]
    \centering
    \begin{minipage}[b]{0.49\textwidth}
        \centering
        \includegraphics[width=\textwidth]{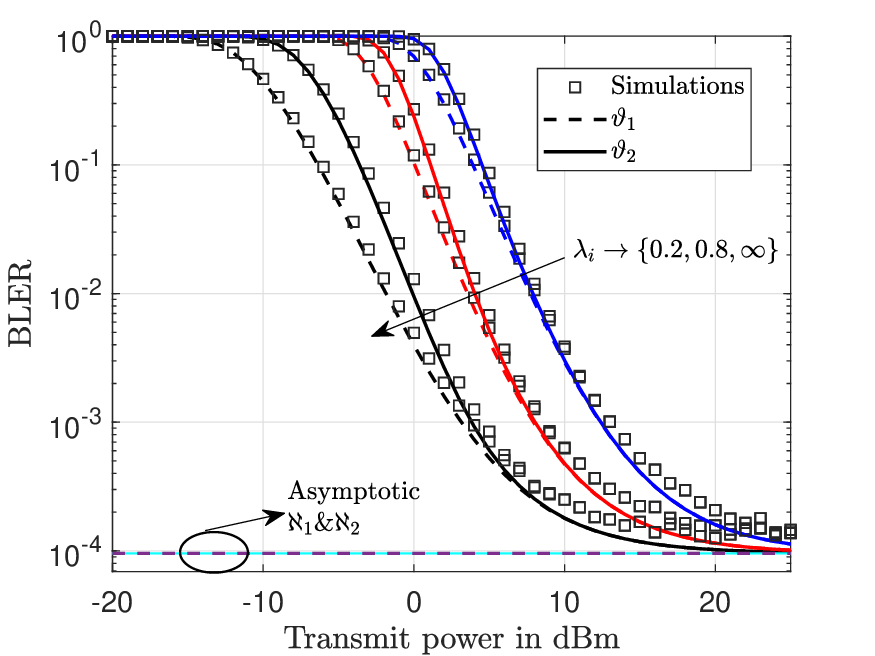}
        \caption{Effect of imperfect CSIR on BLER.}
        \label{fig:ICSI}
    \end{minipage}
     \begin{minipage}[b]{0.49\textwidth}
        \centering
        \includegraphics[width=\textwidth]{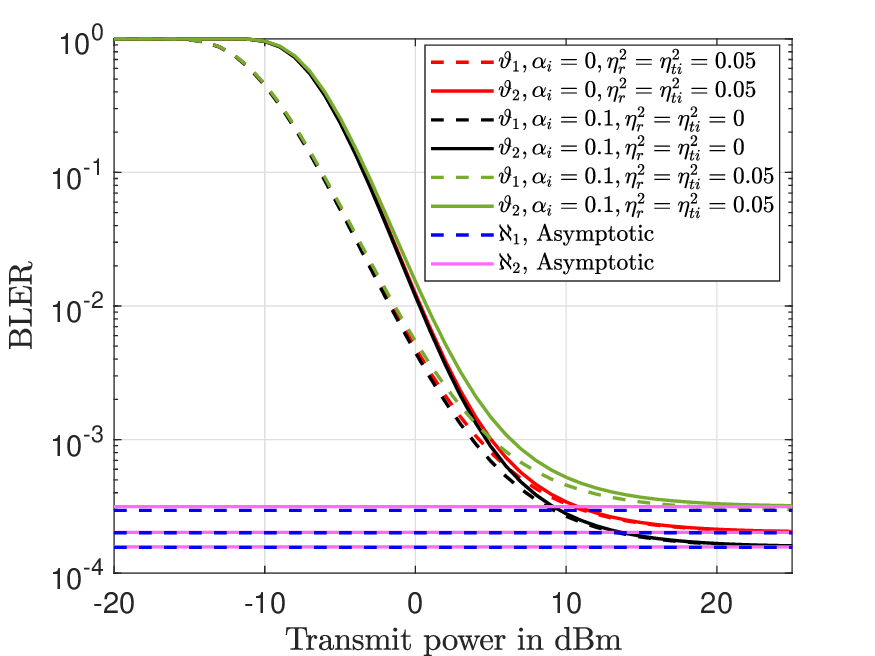}
        \caption{Effect of imperfect SIC and hardware impairments.}
        \label{fig:SIC_HI}
    \end{minipage}
\end{figure*}

Fig.~\ref{fig:ICSI} illustrates the BLER performance of $U_1$ and $U_2$ versus transmit power ($P_i$) under varying channel estimation quality levels ($\lambda_i$), assuming perfect SIC and ideal hardware impairments. The close alignment between simulated and analytical results confirms the accuracy of the BLER expressions derived in \eqref{BLER_U_1} and \eqref{BLER_U_2}.
Notably, the impact of imperfect CSIR is more pronounced at lower transmit power levels. For example, at $P_i = 0$~dBm, both users experience a BLER close to 1 when $\lambda_i = 0.2$. As expected, the performance improves significantly as $\lambda_i$ increases, with the best performance observed when $\lambda_i \to \infty$. However, at higher transmit power levels (i.e., $P_i > 20$~dBm), all BLER curves tend to saturate around approximately $10^{-4}$ for any $\lambda_i > 0$.


Furthermore, the results suggest that the adverse effects of imperfect CSIR can be offset by increasing the transmit power, especially in the high-SNR regime.
 This trend is consistent with the asymptotic BLER behavior analyzed when $\rho_i \to \infty$ as (i.e., as $P_i\to\infty$) and $\lambda_i > 0$, which further supports the validity of the derived analytical expressions in \eqref{Asymp_BLER_U_1} and \eqref{Asymp_BLER_U2}. 

Fig.~\ref{fig:SIC_HI} presents the BLER and asymptotic BLER of $U_1$ and $U_2$ under varying transmit power levels, considering different levels of imperfect SIC ($\alpha_i$) and hardware impairments ($\eta^2_{r}, \eta^2_{ti}$), with perfect CSIR assumed ($\lambda_i \to \infty$). It can be observed that hardware impairments have a more detrimental impact on BLER performance compared to imperfect SIC.  For instance, when $\eta^2_{r} = \eta^2_{ti} = 0.05$ and $\alpha_i = 0$, the BLER for both users saturates at approximately $2 \times 10^{-4}$ for $P_i \approx 20$~dBm. In contrast, with ideal hardware ($\eta^2_{r} = \eta^2_{ti} = 0$) and imperfect SIC $\alpha_i = 0.1$, the BLER is lower, around $1 \times 10^{-4}$ for the same transmit power.
This behavior arises from two interrelated factors. First, unlike hardware impairment terms which appear in all SINR expressions imperfect SIC only affects the decoding process after the stream in \eqref{U_22_SINR} has been decoded. Second, the SIC imperfection term is scaled by a portion of $P_2$ due to message splitting, whereas hardware impairment terms are applied to the entire transmit power, leading to a more pronounced impact on overall performance.

Moreover, a comparative assessment of Fig.~\ref{fig:ICSI} and Fig.~\ref{fig:SIC_HI} reveals an important trend: at low transmit power levels, imperfect CSIR dominates BLER degradation, followed by hardware impairments and then imperfect SIC. However, at high transmit power levels, the influence of imperfect CSIR diminishes, and hardware impairments become the dominant limiting factor, followed by imperfect SIC.
These insights underscore a critical design implication: ignoring fundamental imperfections such as channel estimation errors, hardware impairments, or imperfect SIC when evaluating possible technologies for 6G network architectures can lead to significantly overestimated theoretical performance projections. 

\begin{figure*}[t!]
    \centering
    \begin{minipage}[b]{0.49\textwidth}
        \centering
        \includegraphics[width=\textwidth]{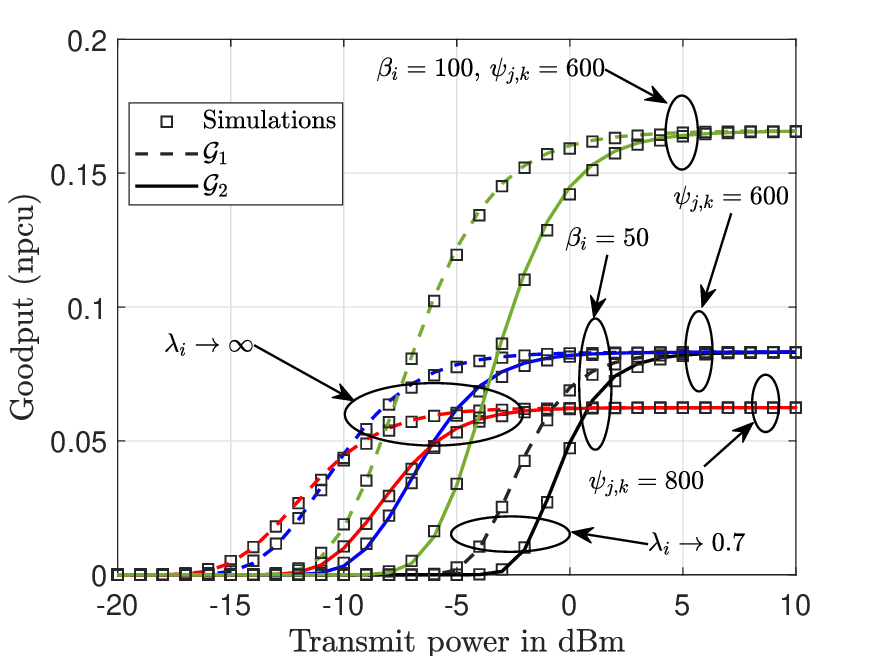}
        \caption{Goodput Versus Transmit Power.}
        \label{goodput}
    \end{minipage}
     \begin{minipage}[b]{0.49\textwidth}
        \centering
        \includegraphics[width=\textwidth]{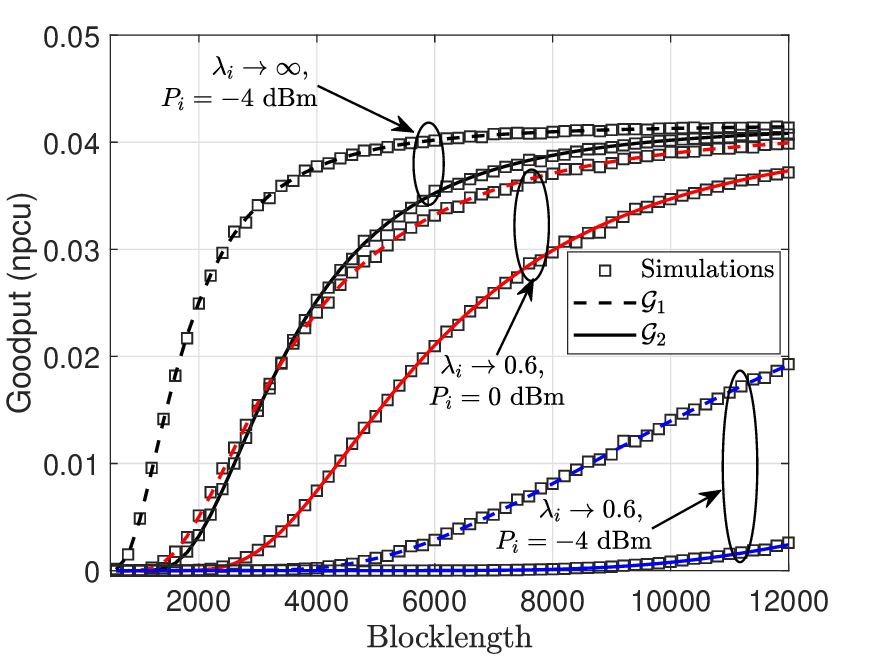}
        \caption{Goodput versus Blocklength.}
        \label{blocklength}
    \end{minipage}
\end{figure*}

Fig.~\ref{goodput} illustrates the goodput performance $\mathcal{G}_1$ and $\mathcal{G}_2$, measured in npcu for both the users, across varying transmit power for different blocklengths $\left(\psi_j\; \forall  j\in\{21, 1\}\right)$, $\left(\psi_k\; \forall k \in\{21, 1, 22\}\right)$, bits per slot values ($\beta_i \forall\; i\in\{1,2\}$) and $\lambda_i$ values with $\alpha_i=0.1$ and $\eta^2_{r} = \eta^2_{ti} = 0.05$. As expected, both $\mathcal{G}_1$ and $\mathcal{G}_2$ increase with rising $\rho_i$, eventually saturating after certain transmit power values. This saturation occurs because the target rates ($r_i$) are fully met, and additional power no longer contributes to meaningful gains due to target rate and rate allocation factor constraints.
Furthermore, the figure shows that at a fixed blocklength of $600$, increasing the bits per slot to $100$ for $U_1$ and $U_2$ significantly boosts their goodput. This is attributed to the increase in transmission rate, which despite slightly higher BLER results in greater overall data throughput. In contrast, when the blocklength is increased from $600$ to $800$ while keeping the bits per slot fixed at $50$, the goodput decreases. Although a longer blocklength improves reliability by lowering BLER, it also reduces the transmission rate, leading to fewer bits transmitted per channel use. As a result, the goodput declines.
These observations highlight the trade-off between rate and reliability under FBL constraints and underscore the importance of carefully balancing system parameters when designing HRLLC systems. Additionally, the expressions derived in \eqref{goodput_U1} and \eqref{goodput_U2} closely match the simulation results, thereby validating the accuracy of the analytical expressions.

  \begin{figure}[t]
    \centering
    \begin{minipage}[b]{0.49\textwidth}
        \centering
        \includegraphics[width=\textwidth]{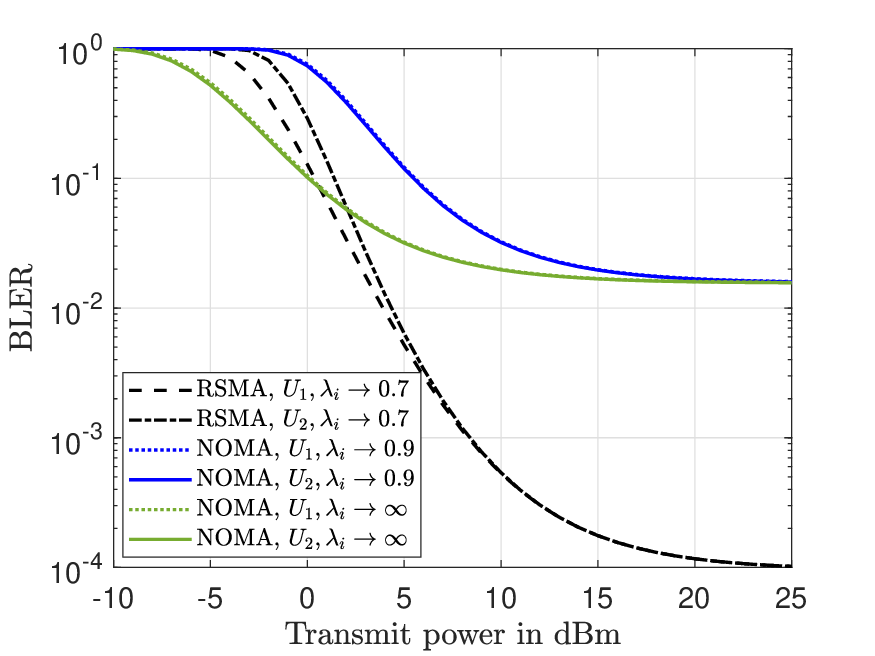}
        \caption{RSMA versus NOMA.}
        \label{NOMA}
    \end{minipage}
    \vspace{-2em}
\end{figure}

Fig.~\ref{blocklength} illustrates the goodput performance of $U_1$ and $U_2$ as a function of blocklength, under different values of the channel estimation quality parameter $\lambda_i$. The number of bits per slot is fixed at $500$, with imperfect SIC ($\alpha_i = 0.1$) and hardware impairments ($\eta^2_{r} = \eta^2_{ti} = 0.05$) taken into account. As blocklength increases, both $\mathcal{G}_1$ and $\mathcal{G}_2$ show a consistent improvement, driven by the enhanced error correction capabilities afforded by longer transmission frames.
However, this upward trend begins to saturate beyond a blocklength of approximately $11000$, where saturation is observed particularly for $\lambda_i \to \infty$ at $P_i = -4$~dBm and for $\lambda_i \to 0.6$ at $P_i = 0$~dBm. This saturation reflects a fundamental trade-off, also observed in Fig.~\ref{goodput}: although longer blocklengths reduce BLER, the fixed bits/slot constraint limits the achievable rate. Consequently, the increased reliability gained from longer frames no longer translates into meaningful gains in goodput, as the extended frame durations start to dominate and reduce goodput efficiency.
This behavior highlights a critical design insight: there exists a practical threshold on blocklength beyond which further increases lead to diminishing returns in terms of goodput. Rather than enhancing performance, excessively long frames can introduce latency and resource inefficiency, ultimately constraining the system's ability to leverage reliability improvements for higher data rates.

Furthermore, Fig.~\ref{blocklength} underscores the significant impact of imperfect CSIR, especially at low transmit power levels. For instance, when $\lambda_i \to 0.6$ and $P_i = -4$~dBm, both $\mathcal{G}_1$ and $\mathcal{G}_2$ remain near zero until a blocklength of approximately $4000$. This observation reinforces the importance of precise channel estimation for maintaining acceptable goodput under constrained transmit power and latency budgets. It also cautions against relying on idealized assumptions, as doing so may mask critical performance limitations in practical HRLLC scenarios.

 Fig.~\ref{NOMA} compares the BLER performance of uplink RSMA and uplink NOMA schemes as a function of transmit power, under the assumptions of imperfect CSIR, perfect SIC, and ideal hardware impairments. For a fair comparison, the NOMA scenario uses 100 bits/slot and a blocklength of 600 for $U_2$, aligning with the RSMA configuration where 50 bits/slot are allocated to each of $U_2$’s split messages ($\chi_{21}$ and $\chi_{22}$) totaling 100 bits/slot.
A key distinction lies in the transmission strategy: unlike RSMA, which splits messages into two streams for layered decoding, NOMA transmits a single message per user. This leads to a higher effective rate per channel use in NOMA but also makes it more sensitive to channel imperfections.

The figure clearly shows that RSMA maintains lower BLER for both users compared to NOMA  even under harsher CSIR conditions. For instance, RSMA users with $\lambda_i \to 0.7$ outperform NOMA users even when the latter has a better channel estimation quality of $\lambda_i \to 0.9$. This highlights RSMA’s greater robustness to CSIR imperfections in uplink transmissions. Moreover, NOMA exhibits early saturation in BLER performance, whereas RSMA continues to show improvement at higher transmit power levels before saturating, underscoring its resilience in maintaining performance over a wider operating range.
Another important observation is the disparity in user dependency. In RSMA, $U_1$ can still decode its message even if $U_2$ fails to decode the second part of its message (as long as $\chi_{21}$ is successfully decoded, as per \eqref{U_22_SINR} and \eqref{U_1_SINR}). In contrast, NOMA’s strict decoding hierarchy means $U_1$ is fully dependent on $U_2$'s successful decoding, making it more vulnerable to errors in shared decoding paths.
These findings collectively demonstrate RSMA’s superior resiliency and flexibility in uplink scenarios, especially under non-ideal conditions. The ability to mitigate inter-user dependency and manage imperfect CSIR more effectively positions RSMA as a more robust solution for uplink multiple access in next-generation networks. It is worth noting that NOMA cannot be obtained by simply setting the power of one of $U_2$'s split sub-messages to zero within the RSMA framework, as this would leave the corresponding rate requirement $r_{22}$ nonzero while the achievable capacity collapses to zero, yielding $\vartheta_{22}=1$ across the entire power range. Instead, the NOMA benchmark is obtained by removing the message split altogether, encoding $U_2$'s full message as a single stream and rate $r_2=\beta_2/\psi_2$.

\vspace{1em}
\textit{Design implications for practical IoT scenarios:}

The numerical results indicate that the design of uplink RSMA-based IoT systems is governed by interdependent trade-offs that vary with the operating regime. In particular, the dominant performance bottleneck is not fixed: at low transmit power levels, system performance is primarily limited by channel estimation accuracy, whereas at higher power levels hardware impairments become the main limiting factor. This regime-dependent behavior implies that improving any single component in isolation does not necessarily translate into overall system performance gains, since the dominant impairment shifts with operating conditions.

Second, the uplink RSMA-specific control parameters, namely the rate allocation factor and power splitting factor, exhibit strong sensitivity to system configuration. The results show that BLER performance degrades significantly outside a relatively narrow operating range, indicating that improper parameter selection can lead to substantial performance loss and increased inter-user dependency. This suggests that neither fixed parameter settings nor offline optimized (or near-optimal) tuning may provide robust performance across varying operating conditions in practical uplink deployments, particularly in dynamic IoT environments.

Third, although increasing blocklength improves reliability by reducing BLER, it simultaneously reduces the effective rate per channel use under a fixed payload constraint. As a result, the gains in reliability eventually saturate and are no longer sufficient to offset the reduced rate, leading to diminishing or even negative improvements in goodput. Overall, the results suggest that practical uplink RSMA-enabled IoT systems may benefit from adaptive configuration of transmission parameters and RSMA allocation factors according to channel conditions and QoS requirements, rather than relying on static parameter settings optimized for a single operating point.

\section{Conclusion}
This work presented a comprehensive performance analysis of uplink RSMA in the FBL regime, explicitly incorporating key practical constraints such as imperfect CSIR, imperfect SIC, and hardware impairments. Closed-form expressions for BLER, goodput and asymptotic BLER are derived. Our findings highlight several important insights for the design of uplink RSMA in HRLLC-oriented systems. First, we observed that at low transmit power levels, imperfect CSIR causes severe performance degradation, while at higher transmit power, the impact of hardware impairments and SIC errors becomes more pronounced. The analysis also revealed a non-trivial trade-off between reliability and goodput. We further explored the sensitivity of RSMA performance to rate allocation and power allocation strategies, providing practical guidance for RSMA parameter configuration under different reliability requirements. Our results demonstrated that careful tuning of the rate allocation factor is essential to maintain balanced user reliability and fairness. Additionally, the power allocation factor for $U_2$’s message split plays a significant role in user interdependence and BLER performance. In comparison to NOMA, RSMA consistently outperformed under realistic conditions, demonstrating stronger resilience to CSIR imperfections and less inter-user dependency. These findings provide practical design guidance for future 6G deployments requiring reliable short-packet communication, particularly dense industrial IoT and mMTC scenarios with imperfect CSIR, SIC, and hardware impairments. The developed analytical framework can further support the design and configuration of uplink RSMA-based access for emerging 6G applications where reliability, latency, and resource efficiency must be jointly considered.


\appendices
 \section{}
\begin{proof}
    To derive the BLER for $U_1$, we first need to evaluate the BLER for $U_2$'s first stream, $x_{21}$. The BLER at the AP for the $s_{21}$ stream can be expressed as:
    \begin{align}\label{Qfunc_U_x21}
        \vartheta_{21} \approx \mathbb{E}\left\{Q\left(\frac{C\left( \Upsilon_{21}\right)-r_{21}}{\sqrt{V\left( \Upsilon_{21}\right) / \psi_2}}\right)\right\}.
    \end{align}
    By employing a piecewise linear approximation, we can simplify \eqref{Qfunc_U_x21} as $Q\left(\frac{C\left( \Upsilon_{21}\right)-r_{21}}{\sqrt{V\left( \Upsilon_{21}\right) / \psi_{2}}}\right) \simeq \Theta_{21}\left({\Upsilon_{21}}\right)$, which is provided in \eqref{qexpand} on the top of next page. Subsequently, by taking the mean of $\Theta_{21}\left({\Upsilon_{21}}\right)$, characterized by the PDF of $\Upsilon_{21}$, and applying integration by parts, we simplify \eqref{qexpand} as follows:
\begin{align}\label{partsintegral}
    \vartheta_{21} \approx \Lambda_{21}\int\limits^{\zeta_{21}}_{\xi_{21}}F{_{\Upsilon_{21}}}(z) dz.
\end{align}
Next, solving the integral in \eqref{partsintegral} by using the Riemann integral approach, $\int\limits^{\zeta_{21}}_{\xi_{21}} g(z)dz \approx ({\zeta_{21}}-{\xi_{21}})g\left(\frac{{\zeta_{21}}+{\xi_{21}}}{2}\right)$, we arrive at the following result:
\begin{align}\label{Riemann}
     \vartheta_{21} \approx \Lambda_{21}\left({\zeta_{21}}-{\xi_{21}}\right)F{_{\Upsilon_{21}}}\left(\frac{{\zeta_{21}}+{\xi_{21}}}{2}\right),
\end{align} where $F{_{\Upsilon_{21}}}({\cdot})$  represents the CDF as expressed in \eqref{U_22_SINR}. To evaluate \eqref{Riemann}, we begin by determining the CDF of ${\Upsilon_{21}}$ expressed as
\begin{align}\label{CDF_21}
F_{\Upsilon_{21}}(z) = \mathrm{Pr}\left(\Upsilon_{21} <z\right).
\end{align}
By substituting \eqref{U_22_SINR} in \eqref{CDF_21} and performing some algebraic manipulations, we obtain
\begin{align}\label{CDF_step21}
    F_{\Upsilon_{21}}(z) =\mathrm{Pr}\!\left(\left|\hat{h}_2\right|^2< \frac{z\left(\left|\hat{h}_1\right|^2B_{11}+C_{21}\right)}{\rho_2\left(\varkappa_{21}-zA_{21}\right)}\right),
\end{align}
where $A_{21}\! =\!\left(\varkappa_{22}+\eta^2_{r}+\eta^2_{t2}\right) $, $B_{11}=\rho_1\left(1 + \eta^2_{t1} +\eta^2_{r}\right)$, $C_{21}=\left(\rho_2\Omega_{g2e}\left(1 + \eta^2_{t2} +\eta^2_{r}\right)+B_{11}\Omega_{g1e}+1\right)$. Let $\mathcal{Z}=|\hat{h}_2|^2$ and $Y=|\hat{h}_1|^2$, then \eqref{CDF_step21} can be expressed as  
\begin{align}\label{Proof_21}
              F_{\Upsilon_{21}}(z)\!=\! \int\limits^{\infty}_{0}\!\!f_{Y}(y) F_{\mathcal{Z}}\!\!\left(\frac{z\left(yB_{11}+C_{21}\right)}{\rho_2\left(\varkappa_{21}-zA_{21}\right)}\right)dy.
         \end{align}
         By substituting \eqref{CDF} in \eqref{Proof_21}, we get
         \begin{align}\label{Proof_22}
              F_{\Upsilon_{21}}(z)\!\!=\!\! \int\limits^{\infty}_{0}\!\!f_{Y}(y)\frac{1}{\Gamma(m_2)}\gamma\left(m_2, {{D}_{21}\left(y B_{11}+C_{21}\right)}\right)dy.
         \end{align}
         As $m_2$ is a positive integer, the gamma distribution simplifies to an Erlang distribution, leading to the expansion of \eqref{Proof_22} as
         \begin{align}\label{erlag}
              F_{\Upsilon_{21}}(z)\!\!=\!\!&\int\limits^{\infty}_{0}f_{Y}(y)\big[1-\exp\left(-{D}_{21}\left(y B_{11}+C_{21}\right)\right)\nonumber\\
    &\times\sum\limits_{{\mathcal{I}}=0}^{{m_2}-1}\frac{\left(D_{21}\left(y B_{11}+C_{21}\right)\right)^{\mathcal{I}}}{{\mathcal{I}}!}\big]dy.
         \end{align}
         Next, by substituting the PDF from \eqref{pdf} into \eqref{erlag} and applying the formula from \cite[$1.111$]{2015249} to the term ${\left({D}_{21}\left(y B_{11}+C_{21}\right)\right)^{\mathcal{I}}}$, we obtain
         \begin{align}\label{erlang_expanded}
              &F_{\Upsilon_{21}}(z)= 1-\sum\limits_{{\mathcal{I}}=0}^{{m_2}-1}\sum\limits_{{\mathcal{J}}=0}^{{\mathcal{I}}}\binom{{\mathcal{I}}}{{\mathcal{J}}}\left({D}_{21}B_{11}\right)^{\mathcal{J}}  \left({D}_{21}C_{21}\right)^{({\mathcal{I}}-{\mathcal{J}})}\nonumber\\
     &\times L_{21}^{m_1}\frac{\exp\left(-{D}_{21}C_{21}\right)  }{({\mathcal{I}}!)\Gamma(m_1)}  \nonumber\\
     &\times\int\limits^{\infty}_{0} y^{{m_1}-1+{\mathcal{J}}}\exp\left(-y\left({D}_{21}B_{11}+L_{21}\right)\right)dy,
         \end{align}
         where $L_{21}=\frac{m_1}{\hat{\Omega}_1}$. Using \cite[$3.381.4$]{2015249}, the integral in \eqref{erlang_expanded} can be solved. By substituting the resulting expression in \eqref{Riemann}, we obtain \eqref{BLER_21}. Next, we solve for $s_1$ stream whose expression can be given as
    \begin{align}\label{Qfunc_U_1}
        \vartheta_{11} \approx \mathbb{E}\left\{Q\left(\frac{C\left( \Upsilon_{1}\right)-r_{1}}{\sqrt{V\left( \Upsilon_{1}\right) / \psi_1}}\right)\right\}.
    \end{align}
    By following a similar approach from \eqref{partsintegral} to \eqref{CDF_21} and after mathematical simplifications, we get
    \begin{align}\label{CDF_step11}
    F_{\Upsilon_{1}}(x) =\mathrm{Pr}\!\left(\left|\hat{h}_1\right|^2< \frac{x\left(yZ_{1}+Z_{2}\right)}{\rho_1\left(1-E_1 x\right)}\right),
\end{align}
where $E_{1}\! =\!\left(\eta^2_{t1} +\eta^2_{r}\right) $, $Z_1=\left(\Xi_2+\eta^2_{t2} +\eta^2_{r}\right)\rho_2$ and $Z_{2}=\left(\rho_1\Omega_{g1e}\left(1+\eta^2_{r}+\eta^2_{t1}\right)+\Omega_{g2e}Z_1 +1\right)$.  Let $X=|\hat{h}_1|^2$ and $Y=|\hat{h}_2|^2$, then \eqref{CDF_step11} can be expressed as  
\begin{align}\label{Proof_11}
              F_{\Upsilon_{1}}(x)\!=\! \int\limits^{\infty}_{0}\!\!f_{Y}(y) F_{X}\!\!\left(\frac{x\left(yZ_{1}+Z_{2}\right)}{\rho_1\left(1-xE_{1}\right)}\right)dy.
         \end{align}
         By substituting \eqref{CDF} in \eqref{Proof_11}, we get
         \begin{align}\label{Proof_1x1}
              F_{\Upsilon_{1}}(x)\!\!=\!\! \int\limits^{\infty}_{0}\!\!f_{Y}(y)\frac{1}{\Gamma(m_1)}\gamma\left(m_1, {W_{1}\left(y Z_{1}+Z_{2}\right)}\right)dy,
         \end{align}
         where  $W_{1}\!=\!\frac{m_1 \left(\left(\zeta_{11}+\xi_{11}\right)/2\right) }{{\hat{\Omega}_1}\rho_1\left(1-\!\left(\left(\zeta_{11}+\xi_{11}\right)/2\right)E_1\right)}$. Now, by following the same methodology as outlined from \eqref{erlag} to \eqref{erlang_expanded}, the final expression for $\vartheta_{11}$ is obtained in \eqref{BLER_11}.
\end{proof}
\begin{figure*}
 \begin{equation}\label{qexpand}
\Theta_{21}(\Upsilon_{21})\approx\begin{cases}
1, &    {\Upsilon_{21}}\leq \xi_{21}, \\
0.5- \Lambda_{21}\left({\Upsilon_{21}}-\mu_{21}\right), & \xi_{21}\leq  {\Upsilon_{21}} \leq \zeta_{21}\\
0, & {\Upsilon_{21}}\geq  \zeta_{21}, \end{cases}
\end{equation}\hrulefill
\end{figure*}
\section{}
  \begin{proof}
Since no imperfections are assumed in the BLER equation provided in Remark~2, both $\vartheta_{21}$ and $\vartheta_1$ can be easily derived. However, the expression for $\Upsilon_{22}$ in \eqref{U_22_final_SINR} changes and can be rewritten as $\Upsilon_{22} = \frac{|h_2|^2 \rho_2 \varkappa_{22}}{1}$. Next, the CDF of $\Upsilon_{22}$ can be written as
\begin{align}
    F_{\Upsilon_{22}}(x) = \mathrm{Pr}\left( |h_2|^2 < \frac{x}{\rho_2 \varkappa_{22}} \right).
\end{align}
Finally, by substituting the expression from \eqref{CDF} into the equation above, we obtain $\vartheta_{22}$.
\end{proof}

\section{}
\begin{proof}
 To derive the asymptotic BLER expression for $U_1$, we begin by following the same steps described from \eqref{Qfunc_U_x21} to \eqref{Proof_21}. Due to the channel estimation error variance formula used, the term $Z_{2}$ simplifies to 1 as $\rho_1 = \rho_2 \to \infty$. 
Next, by applying \textit{L'Hôpital's rule}, we obtain the following expression:
\begin{align}\label{asymp_proof}
    \mathfrak{T}_{21} = \int\limits^{\infty}_{0}\!\!f_{Y}(y)\left[1 - \exp\left(-\mathfrak{D}_{2,1}y\right)\sum\limits_{{\mathcal{I}}=0}^{m_2-1} \frac{\left(\mathfrak{D}_{2,1} y\right)^{\mathcal{I}}}{{\mathcal{I}}!}\right]dy,
\end{align}
      where $\mathfrak{D}_{21}=\frac{m_2\hat{B}_{11}x}{\hat{\Omega}_2\left(\varkappa_{21}-xA_{21}\right)}$ and $\hat{B}_{11}=\left(1+\eta^2_{t1}+\eta^2_{r}\right)$. Following a similar solution approach as presented in Appendix~A, the final expression for $\mathfrak{T}_{21}$ is obtained. Similarly, the asymptotic BLER corresponding to \eqref{U_1_SINR} can also be obtained. Using both $\mathfrak{T}_{21}$ and $\mathfrak{T}_{11}$, we obtain \eqref{Asymp_BLER_U_1}.
     \end{proof}
     
     \section{}
     \begin{proof}
          For evaluating the  asymptotic BLER of $U_2$  considering  perfect SIC and ideal hardware impairments, we can write the CDF expression as 
          \begin{align}\label{final_asymp}
              F^{\infty}_{\Upsilon_{22}}(x) =  \mathrm{Pr}\left( |h_2|^2 < \frac{x E_{25}}{\rho_2 \varkappa_{22}} \right).
          \end{align}
          As $\rho_1=\rho_2\to\infty$, the term $E_{25}$ is reduced to $1$. Next,  we use $\gamma\left(c,a\right) \underset{a \to 0}{\approx} \frac{a^c}{c}$, thus, we obtain
          \begin{align}
               F^{\infty}_{\Upsilon_{22}}(x) = \frac{1}{\Gamma({m_2})\; m_2}\left({ \frac{m_2\left(\left(\zeta_{22}+\xi_{22}\right)/2\right)}{\hat{\Omega}_2\rho_2\varkappa_{22}}}\right)^{m_2}.
          \end{align}
           By following a similar approach outlined in \eqref{Riemann} for $U_2$'s second stream, we get $\mathfrak{T}_{22}$.
     \end{proof}
\bibliographystyle{IEEEtran_renamed}
	\bibliography{referencing}


\end{document}